\documentclass{PoS}

\usepackage{latexsym,amssymb,amsfonts,amsmath}

\newcommand{\diag}{{\rm diag}}

\newcommand{\la}{\langle}
\newcommand{\ra}{\rangle}

\newcommand{\f}[2]{\frac{#1}{#2}}

\newcommand{\Oc}{{\cal O}}

\newcounter{dumbcount}

\title{Localisation, chiral symmetry and confinement in QCD and
  related theories}

\ShortTitle{Localisation, chiral symmetry and confinement}

\author{\speaker{Matteo Giordano}
\\
ELTE E\"otv\"os Lor\'and University, Institute for Theoretical
Physics, P\'azm\'any P.\ s.\ 1/A, H-1117, Budapest, Hungary, and \\
MTA-ELTE Lend\"ulet Lattice Gauge Theory Research Group, P\'azm\'any
P.\ s.\ 1/A, H-1117, Budapest, Hungary
\\
E-mail: \email{giordano@bodri.elte.hu}}

\abstract{I discuss recent results on the relation between the
  localisation of low-lying Dirac eigenmodes, the restoration
  of chiral symmetry, and deconfinement in QCD and QCD-like
  models, providing evidence of a close connection between the
  three phenomena.}

\FullConference{XIII Quark Confinement and the Hadron Spectrum -
  Confinement2018\\ 
		31 July - 6 August 2018\\
		Maynooth University, Ireland}

\begin{document}

\section{Introduction}
\label{sec:intro}

It is well known that QCD at finite temperature undergoes a
deconfining and chirally-restoring transition, or, more precisely,
that QCD displays an analytic crossover in the range $T\simeq
145$--165 MeV in which both confining and chiral properties change
radically. The pseudocritical temperatures for the chiral and
deconfining transitions, defined from the position of the peaks of the
chiral susceptibility and of the quark entropy, respectively, are
equal within errors~\cite{BW,Bazetal}. For the sake of definiteness we
will take $T_c\simeq 155$ MeV. The close relation between these two
very different phenomena is not unique to QCD, but is found also in
other QCD-like theories, like, e.g., SU(3) pure gauge theory in
3+1~\cite{su3} and 2+1 dimensions~\cite{DHKM}, and $N_f=3$ QCD with 
unimproved rooted staggered fermions on $N_T=4$
lattices~\cite{unimproved1}. The nature of this relation is however
not fully understood yet.

It is also well known that chiral symmetry breaking originates from
the accumulation of eigenvalues of the Dirac operator near the
origin~\cite{BC}: the spectral density is finite at the origin below $T_c$,
while it vanishes above it. The question is then if such an
accumulation, or the lack thereof, is related to the confining
properties of the theory. A possible link, or at least a tool to study
the relation between deconfinement and chiral symmetry restoration, is
a third phenomenon that also takes place at $T_c$, namely the
localisation of the lowest Dirac modes. 

\begin{figure}[t]
  \centering
  \includegraphics[width=.575\textwidth]{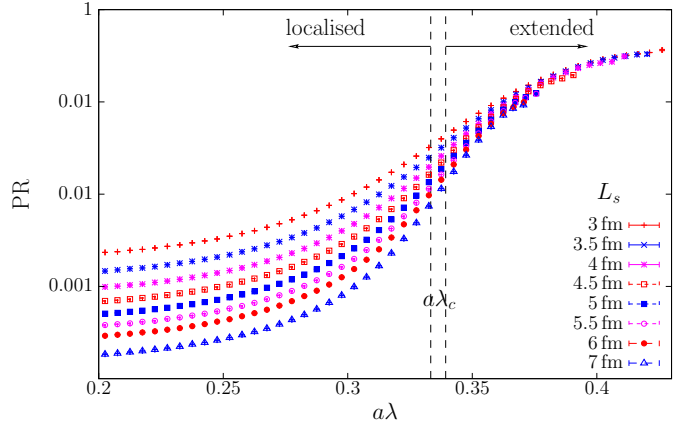}
  \caption{Participation ratio of the low Dirac modes in
    high-temperature QCD (here $T\simeq 400~{\rm MeV}$), obtained on
    the lattice with staggered fermions. Data from
    Ref.~\protect{\cite{crit}}.}  
  \label{fig:pr}
\end{figure}

It is by now well established that while below $T_c$ all the Dirac
modes are extended throughout the whole system, above $T_c$ the low
modes are localised, up to a temperature-dependent critical point
$\lambda_c(T)$ in the spectrum where a localisation/delocalisation
transition, or Anderson transition, takes
place~\cite{GGO,KGT,KP,KP2,crit,Cossu,Holicki:2018sms}. The simplest way to
observe localisation is to measure the participation ratio (PR) of the 
eigenmodes, which measures the fraction of spacetime effectively
occupied by a mode. For normalised eigenvectors $\psi(x)$ of the Dirac 
operator, computed on a hypercubic lattice of spatial volume $V=N_s^3$
and temporal size $N_T$ (in lattice units), one has
\begin{equation}
  \label{eq:prdef}
  {\rm PR} \equiv \f{1}{V N_T}\left[\sum_x
    |\psi^\dag(x)\psi(x)|^2\right]^{-1}\,, 
\end{equation}
where $\psi^\dag(x)\psi(x)$ denotes the scalar product in colour
and Dirac space. As the volume of the system is increased, the PR
remains constant for an extended mode, while it goes to zero for a
mode localised in a finite region. Localisation of the low modes in 
high-temperature QCD can be seen in Fig.~\ref{fig:pr}.

In this contribution, after a brief discussion of 
the main tools required in the study of localisation, I will review
what is known about localisation in QCD. I will then discuss a
possible mechanism for localisation, and how this mechanism relates to
the confining and chiral properties of QCD. I will then proceed to
discuss localisation in QCD-like theories in 3+1 dimensions, briefly
discuss the possible role of topology, and show some preliminary
results for QCD-like theories in 2+1 dimensions.

\section{Numerical studies of localisation}
\label{sec:numloc}

The main tools required to study localisation in QCD are the theory of
disordered Hamiltonians, Lattice Gauge Theory, and Random Matrix
Theory, which I now briefly review.

\subsection{Disordered Hamiltonians}
\label{sec:AM}

Disordered Hamiltonians aim at describing systems containing some form
of disorder. The most famous example is the Hamiltonian of the
Anderson model (AM) for electrons in ``dirty''
conductors~\cite{Anderson}, which consists of the usual tight-binding 
Hamiltonian supplemented with a random on-site potential
$\varepsilon_{\vec x}$, mimicking the presence of impurities in the
crystal, 
\begin{equation}
  \label{eq:AM_O}
   H_{\vec x, \vec y}^{{\rm AM}} = \varepsilon_{\vec x} \delta_{\vec x, \vec y} + 
  \sum_{\mu=1}^3 
  (\delta_{\vec x+\hat\mu,\vec y} +
  \delta_{\vec x-\hat\mu,\vec y})\,.
\end{equation}
In the simplest version of the model, the $\varepsilon_{\vec x}$ are
drawn from a uniform distribution on the interval
$\big[-\f{W}{2},+\f{W}{2}\big]$, whose width $W$ measures the amount of  
disorder in the system. While in the absence of disorder ($W=0$) the
electron states are delocalised Bloch waves, as soon as some disorder
is put into the system ($W\neq 0$) localised modes appear at the band
edge, beyond a critical energy $E_c(W)$, called ``mobility edge''. As
the amount of disorder is increased, $E_c(W)$ moves towards
the band center, until a critical disorder $W_c$ is reached: for
$W>W_c$ all modes are localised, and the metal becomes an insulator.  
As one crosses the mobility edge the system undergoes a second-order
phase transition with divergent correlation length $\xi(E) \simeq
|E-E_c|^{-\nu}$, known as Anderson transition. 
The analogue of $E_c(W)$ in QCD is $\lambda_c(T)$, with
temperature playing the role of control parameter of the amount of
disorder in the system. By contrast with the AM, in QCD
all the modes become extended below $T_c$. 

The AM discussed here is the simplest one, with
diagonal disorder only, but other versions exist. One that is
relevant to us is the so-called unitary Anderson model (UAM), which 
includes also off-diagonal disorder in the form of random phases 
in the hopping terms, mimicking the presence of a random magnetic
field,
\begin{equation}
  \label{eq:AM_U}
   H_{\vec x, \vec y}^{{\rm UAM}} = \varepsilon_{\vec x} \delta_{\vec x, \vec y} + 
  \sum_{\mu=1}^3 
  (\delta_{\vec x+\hat\mu,\vec y} +
  \delta_{\vec x-\hat\mu,\vec y}){e^{i\phi_{\vec x, \vec y}}}\,,
\qquad \phi_{\vec y, \vec x} = -\phi_{\vec x, \vec y}\,.
\end{equation}
Here ``unitary'' refers to the symmetry class in the sense of Random
Matrix Theory, discussed below. The AM of Eq.~\eqref{eq:AM_O} belongs
instead to the orthogonal class. The change in symmetry in turn
affects the critical behaviour at the Anderson transition.

\subsection{Lattice Gauge Theory}
\label{sec:LGT}

Most nonperturbative studies of QCD are based on Lattice Gauge Theory 
(LGT), which deals with the gauge theory functional integral in
Euclidean space discretised on a finite lattice, which is typically a
hypercube of spacing $a$ of sizes $N_S$ and $N_T$ in the spatial and
temporal directions, respectively. Fermion fields $\psi(n)$,
$\bar{\psi}(n)$ live on the lattice sites $n$, while gauge fields
$A_\mu(x)$ are replaced by parallel trasporters,  $U_\mu(n) = {\rm
  Pexp}\{ig\int_{an}^{an+a\hat\mu} A_\alpha(x')dx^{\prime\alpha}\}$
living on the lattice edges $(n,n+\hat{\mu})$. Periodic boundary
conditions are usually imposed on the fields in the spatial
directions. At finite temperature the lattice size $1/T = aN_T$ in the
temporal direction is kept fixed in physical units, and appropriate
boundary conditions are imposed on the fields (periodic for bosons and
antiperiodic for fermions). A detailed review of LGT is way beyond the
scope of this paper. There are however two points worth
emphasising. The first important point is that in LGT the gauge theory
functional integral turns in practice into the partition function of a
statistical system,  
\begin{equation}
  \label{eq:LGT}
  Z = \int {\cal D}U \int {\cal D}\psi \int{\cal D}\bar{\psi}
  \,e^{-S_{\rm gauge}[U] - \bar{\psi} (D_{\rm lat}[U] + m)\psi} \,,
\end{equation}
where the integration extends over a large but finite number of
degrees of freedom. The infinite-volume limit $V\to\infty$ and the
continuum limit $a\to 0$ are eventually taken. Here $S_{\rm gauge}[U]$
and $D_{\rm lat}[U]$ are respectively a discretised version of the
Yang-Mills action and of the Dirac operator. The case I will be
focussing on is that of the staggered discretisation, 
\begin{equation}
  \label{eq:LGT2}
  [D_{\rm stag}]_{n,n'} = \textstyle\f{1}{2}\sum_\mu \eta_\mu(n)
  \left[U_\mu(n) \delta_{n+\hat{\mu}.n'} -  
U_{\mu}(n-\hat{\mu})^\dag \delta_{n-\hat{\mu}.n'} \right]\,, \qquad
\eta_\mu(n) = (-1)^{\sum_{\alpha<\mu}n_\alpha}\,.
\end{equation}
The second relevant point here is that the staggered operator is  
just $i$ times the Hamiltonian of a quantum mechanical system with
purely off-diagonal disorder, provided by the gauge links $U_\mu(n)$,
and so the Dirac operator can be studied just like a disordered
Hamiltonian.

\subsection{Random Matrix Theory}
\label{sec:RMT}

Random Matrix Theory (RMT), quite unsurprisingly, studies  matrices
with random entries, examples of which are the AM Hamiltonian and the
lattice Dirac operator. RMT however typically deals with dense
matrices, for which certain statistical properties of   the spectrum
are universal, depending only on the symmetry class of the matrix
ensemble. An important universal property of dense RMT ensembles is
the distribution of the unfolded level spacings, i.e., the distance 
between consecutive eigenvalues divided by the average level spacing
in the relevant spectral region. The unfolded spectrum is defined by
mapping the eigenvalues $\lambda_i$ to $\lambda_i\to x_i
 = \int^{\lambda_i} d\lambda\, \rho(\lambda)$, where
 $\rho(\lambda)\equiv \la \sum_n\delta(\lambda - \lambda_n)\ra$ is the 
spectral density, and $\la\ldots\ra$ denotes ensemble averaging. After
unfolding, the spectral density becomes identically 1 throughout the 
spectrum. The unfolded level spacing distribution (ULSD) $P(s)$ is the
probability distribution of $s_i=x_{i+1}-x_i$ over the matrix
ensemble. This quantity can be computed analytically for the so-called
Gaussian ensembles, and by universality it describes the unfolded
spectrum for all models in a given symmetry class. The ULSD is very
well approximated by the {\it Wigner's surmise}, $P(s)=as^b
e^{-cs^2}$, with class-dependent constants $a,b,c$. By contrast, 
for independently fluctuating eigenvalues (Poisson statistics) one
finds $P(s)=e^{-s}$.

For the sparse matrices we are interested in, it turns out that the
spectral statistics depend on the spectral region, and are
connected to the localisation properties of the eigenmodes: extended
modes obey the appropriate type of RMT statistics given the symmetries
of the model, while localised modes obey Poisson statistics,
fluctuating independently. This connection can be used to determine
precisely the position of the mobility edge $\lambda_c$ by means of a 
finite-size-scaling study of spectral statistics~\cite{SSSLS}. Indeed,
given a dimensionless quantity $\Oc(\lambda,L)$ derived from the ULSD
and computed locally in the spectrum, i.e., in an infinitesimal
neighbourhood of $\lambda$, one expects under the usual scaling
hypothesis that in the vicinity of $\lambda_c$ it depends on the
system size $L$ as 
$\Oc(\lambda,L)=O(\xi(\lambda)/L)=O(|\lambda-\lambda_c|^{-\nu}/  
L)$. Analyticity in a finite volume
then imposes  $\Oc(\lambda,L)=f( (\lambda-\lambda_c)L^{1/\nu})$. This
relation can be used to determine $\lambda_c$ and the critical
exponent $\nu$, as well as the critical value
$\Oc_c=\Oc(\lambda_c,L)$, which is volume-independent, and expected to
be determined by a universal 
critical statistics. A convenient observable turns out to be the
integrated ULSD,
\begin{equation}
  \label{eq:intULSD}
  I_\lambda = \int_0^{s_0} ds\,P(s)\,,
\end{equation}
where $s_0$ is the crossing point of the appropriate Wigner's surmise
and the exponential function. 

\section{Localisation in Lattice QCD}
\label{sec:LQCD}

As already anticipated in the Introduction, while all the Dirac modes
are delocalised in the low-temperature phase of QCD, above $T_c$ the 
low modes are localised for $\lambda<\lambda_c(T)$, with a
second-order phase transition (Anderson transition) at $\lambda_c(T)$.
Localised modes have been observed with different fermion
discretisations (staggered~\cite{GGO,KP2}, overlap~\cite{KGT,Holicki:2018sms},
domain wall~\cite{Cossu}). 
Most quantitative studies have been carried out with staggered
fermions, on which I will mostly focus here. 

\subsection{Numerical results}
\label{sec:num}

\begin{figure}[t]
  \centering
    \includegraphics[width=.47\textwidth]{lcpmvsT.eps}\hspace{\stretch{1}}
    \includegraphics[width=.515\textwidth]{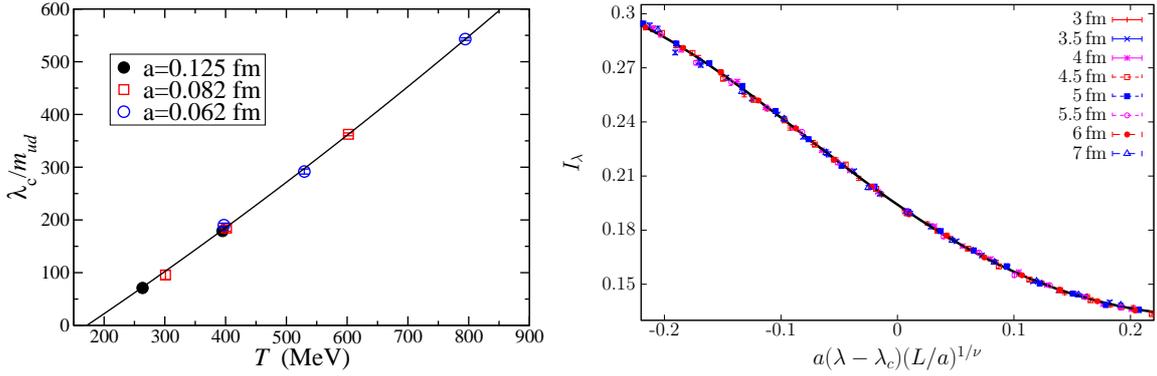}  
  \caption{(Left.) Dependence of the mobility edge on temperature in
    QCD. Figure from Ref.~\protect{\cite{KP2}}.
  (Right.) Scaling function for $I_\lambda$ at the Anderson transition
  in QCD (see text for details). Data from Ref.~\protect{\cite{crit}}.}
  \label{fig:lcvst}
\end{figure}
The dependence of the mobility edge on temperature was studied
in Ref.~\cite{KP2}, using 2+1 flavours of 2-stout improved rooted
staggered fermions with physical quark masses~\cite{BW}. 
As can be seen in Fig.~\ref{fig:lcvst} (left), $\lambda_c(T)$
extrapolates to zero at a temperature compatible with $T_c$. Since it
is a feature of the spectrum, the mobility edge is expected to
renormalise like a quark mass~\cite{GL,KP2}, and so
$\f{\lambda_c}{m_{ud}}$ with $m_{ud}$ the light quark mass is expected
to be the physically relevant, renormalisation-group-invariant
quantity. This ratio seems indeed to be independent of the lattice
spacing, so that localisation is expected to survive the continuum
limit. 

Critical properties at the mobility edge were studied in
Ref.~\cite{crit} with the same type of fermions on $N_T=4$ lattices at
$\beta=3.75$, corresponding to a lattice spacing of $a=0.125~{\rm fm}$
and a temperature of $T=394~{\rm MeV} = 2.6~T_c$. By means of a
finite-size-scaling analysis [see Fig.~\ref{fig:lcvst} (right)] the
critical exponent $\nu$ was found to be $\nu_{\rm QCD} = 1.46(3)$, in
agreement with that obtained for the 3D UAM~\cite{SO}. Further studies
showed that also the multifractal exponents characterising the
eigenvectors at criticality match those of the 3D UAM~\cite{UGPKV}. 

\subsection{Polyakov lines, the ``sea/islands'' picture
  and the Dirac-Anderson Hamiltonian}
\label{sec:PLfluc}

It is a bit surprising that QCD and UAM share the same critical
properties at the Anderson transition. While the unitary symmetry
class is indeed the appropriate one for staggered fermions, and the
dimensionality of high-temperature QCD is indeed expected to be
reduced from 4 to 3, at first sight it is hard to spot any source
of 3D diagonal disorder. The staggered Dirac operator has in fact only 
off-diagonal 4D disorder, for which different critical properties are 
expected. 

The source of diagonal disorder was first identified by the authors of
Ref.~\cite{BKS} in the spatial fluctuations of the Polyakov lines. For
a perfectly ordered configuration with Polyakov lines aligned along
the identity and trivial spatial links, the spectrum displays a sharp
gap corresponding to the lowest Matsubara frequency. For realistic 
configurations the ordering is partial, with a ``sea'' of ordered
Polyakov lines and ``islands'' of fluctuations away from the identity,
so the gap turns into an effective gap, below which the spectral
density is small. Living on the ``islands'' is ``energetically''
favourable, allowing to reduce the eigenvalue and penetrate the gap,
the price to pay being that the mode has to be localised in these
``traps''. The argument was later extended from SU(2) to SU(3) in
Ref.~\cite{GKP}, and has been tested in several ways: on SU(2) 
configurations~\cite{BKS}, in a toy model for QCD~\cite{GKP}, 
and most recently in QCD with domain-wall~\cite{Cossu}
and overlap fermions~\cite{Holicki:2018sms}. 
 
A shortcoming of this argument is that it does not explain why
there is no localisation below $T_c$. The absence of islands
cannot be the answer, since there is no such a thing as islands in the
AM, which displays localised eigenmodes nonetheless. This is simply
due to the presence of fluctuations in the disorder, which certainly
are present in the Polyakov lines also in the low-temperature phase. 
In fact, what one has below $T_c$ is an effectively 4D system
with (relatively weak) 
off-diagonal disorder, as one would expect by taking the Dirac
operator at face value. For this system, no localised modes at the
band center are expected. 

A step in understanding this issue is made by showing that
the staggered operator is formally equivalent to a set of coupled 
3D AMs~\cite{GKP2,GKP3}. This is done by
separating the temporal hoppings, treated as an unperturbed, ``free''
Hamiltonian, from the spatial hoppings, treated as the interaction,
and then diagonalising the free Hamiltonian. After diagonalisation of
its free part, the ``Dirac-Anderson Hamiltonian'' $H=-iD_{\rm stag}$ reads
 \begin{equation}
   \label{eq:DAH2}
   \begin{aligned}
    { H}_{\vec x,\vec y} &=    \delta_{\vec x, \vec y} D(\vec x)
    + \sum_{j=1}^3\f{\eta_j(\vec x)}{2i}\left[
      \delta_{\vec x+\hat\jmath,\vec y} V_{+j}(\vec x)
      -\delta_{\vec x-\hat\jmath,\vec y} V_{-j}(\vec x)
\right]\,,
 \\
     \left[D(\vec x)\right]_{ak,bl} &= 
\eta_4(\vec  x)\sin\omega_{ak}(\vec x)\delta_{ab}\delta_{kl}\,, \quad
  \left[V_{\pm j}(\vec x)\right]_{ak,bl}
=    \f{1}{N_T} \sum_{t=0}^{N_T-1}
    e^{i\f{2\pi t}{N_T}(l-k)}      \left[
\tilde{U}_{\pm j}(t,\vec x)\right]_{ab}\,.
   \end{aligned}
 \end{equation}
Here $\tilde{U}_{\pm j}$ are the spatial links in the 
``uniform temporal diagonal gauge'', or Polyakov gauge, defined by the
condition  
\begin{equation}
  \label{eq:DAH5}
 U_4(t,\vec x)
=[P(\vec x)]^{\f{1}{N_T}} =\diag\left(e^{i\f{\phi_1(\vec
      x)}{N_T}},\ldots,e^{i\f{\phi_{N_c}(\vec x)}{N_T}}\right)\,, 
\end{equation}
where $e^{i\phi_a(\vec x)}$, $a=1,\ldots,N_c$, are the eigenvalues of
the Polyakov line at $\vec{x}$, subject to the condition $\sum_a
\phi_a(\vec x) = 0$. The unperturbed eigenvalues are determined by the
effective Matsubara frequencies,
\begin{equation}
  \label{eq:DAH3}
 \omega_{ak}(\vec x) = {\textstyle\f{1}{N_T}}(      \pi + \phi_a(\vec x)
        + 2\pi k) \,,
\end{equation}
where the indices correspond to the spatial site $\vec x$, colour $a$,
and temporal momentum $k=0,\ldots,N_T-1$. One then finds precisely
$N_T$ 3D AMs with diagonal disorder provided by the Polyakov line
phases through the effective Matsubara frequencies, plus off-diagonal
disorder coming from the interaction. These models are also coupled by
the interaction, and the strength of the coupling is related to the
correlation between time slices: the stronger the correlation, the
weaker the coupling.  

Unlike the simple AMs discussed previously, the strength of the
disorder (both diagonal and off-diagonal) is bounded here (the
matrices $V_{\pm j}(\vec{x})$ can be shown to be unitary), so one
cannot simply induce localisation by increasing the noise. However, in
the two phases of QCD the nature of the disorder is different. In the
disordered phase there are weak correlations among time slices, and no
structure in the diagonal noise, so strong coupling of the AMs and
uncorrelated diagonal noise. In the ordered phase there are strong
correlations among time slices due to ordering of the Polyakov lines,
which also provide a ``sea'' in which ``islands'' of convenient
localising centers are found. This implies weak coupling of the AMs
and correlated diagonal disorder. Studies in a QCD-based toy  
model~\cite{GKP2} show that both the correlations of time slices and
the presence of islands are needed to localise the low modes.

\subsection{Chiral symmetry restoration and localisation from
  deconfinement?} 
\label{sec:decchiloc}

It is now time to ask the important question: how are deconfinement,
chiral symmetry restoration, and localisation of the low modes related?
Although I cannot give a definite answer, I can at least list a few
points that the answer should address.
\begin{enumerate}
\item Deconfinement changes the effective dimensionality of the
    Dirac-Anderson system and creates an effective gap in the
    spectrum.
\end{enumerate}
At low temperature the AMs are strongly coupled,
so the temporal momentum is effectively one more dimension, and the
Dirac-Anderson system is then effectively 4D. This is not surprising,
but then there is nothing to gain in recasting the Dirac operator in
the Dirac-Anderson form, and the purely off-diagonal nature of the
disorder is already captured in the usual form of the staggered
operator. At high temperature instead the coupling is weak, and the
system acts as a collection of 3D AMs with diagonal disorder, weakly
interacting with each other.  
\setcounter{dumbcount}{\theenumi}
\begin{enumerate}
\setcounter{enumi}{\thedumbcount}
\item  Localisation requires the presence of an effective gap and
  decoupling of the AMs. 
\end{enumerate}
In the presence of an effective gap, islands are convenient for
localisation, but if the AMs are too strongly coupled then
the mixing of modes can prevent localisation. This was understood 
studying the toy model of Ref.~\cite{GKP2}. 
\setcounter{dumbcount}{\theenumi}
\begin{enumerate}
\setcounter{enumi}{\thedumbcount}
\item Chiral symmetry breaking, i.e., a finite density of near-zero
    modes, requires that the AMs be coupled, and that a finite density
    of small unperturbed modes be present. 
\end{enumerate}
Also this was understood studying the toy model of Ref.~\cite{GKP2}:
eigenmodes do not accumulate near the origin if there is no mixing of
temporal momenta, and if there are not enough low unperturbed modes.  

The conclusion is that deconfinement seems to precisely provide
the conditions for the localisation of Dirac modes, and to remove the
conditions for chiral symmetry breaking. If the picture above is
correct, this should happen in gauge theories on quite general
grounds, since nothing specific to QCD has been used: all that is
required is the ordering of the Polyakov lines. If so, then the
coincidence of deconfinement, chiral symmetry restoration and
localisation of the low modes should be a rather general
phenomenon. This is what I am going to discuss next.

\section{Localisation in 3+1D QCD-like theories}
\label{sec:qcdlike}

In this section I review results about the connection between
deconfinement, chiral symmetry restoration and localisation in 3+1D
QCD-like theories. To study this connection, theories with a genuine
phase transition are better suited, since they provide a clean-cut
situation with clearly separated phases. 

\begin{figure}[t]
  \centering
  \includegraphics[width=0.5175\textwidth]{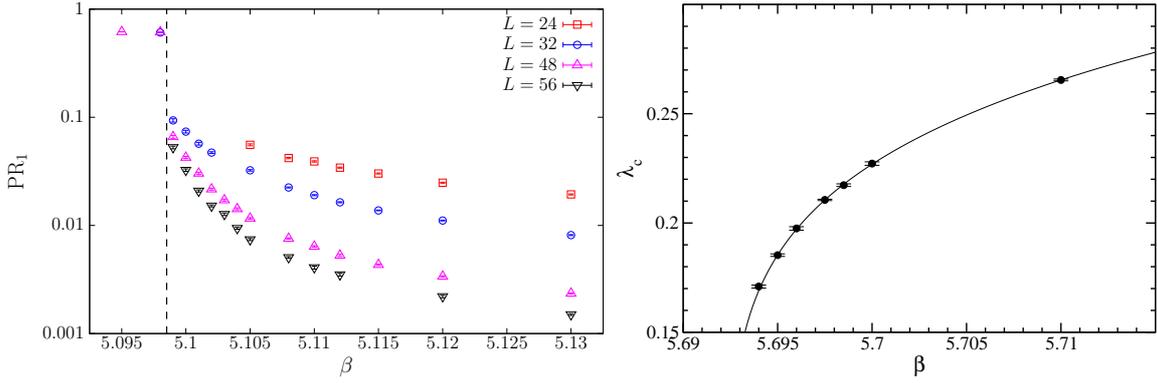}\hspace{\stretch{1}}
  \includegraphics[width=0.472\textwidth]{lc_vs_beta4.eps}
  \caption{(Left.) PR of the lowest mode as a function of the
    coupling for $N_f=3$ unimproved staggered fermions on $N_T=4$
    lattices. Data from Ref.~\protect{\cite{GKKP}}.
    (Right.) Mobility edge as a function of the coupling in SU(3) pure
    gauge theory. Figure from Ref.~\protect{\cite{KoVi}}.}
  \label{fig:nt4us}
\end{figure}
The first case I will discuss is $N_f=3$ QCD with unimproved rooted
staggered quarks on lattices with $N_T=4$, studied in
Ref.~\cite{GKKP}. For light enough quarks this model displays a first
order phase transition at some critical value of the
coupling~\cite{unimproved1}, with both the Polyakov loop expectation
value and the chiral condensate showing a jump. Although this
transition is a lattice artefact that does not survive the continuum
limit, nevertheless this model considered at fixed lattice spacing is
a perfectly good statistical model where we can test our ideas about
deconfinement, chiral restoration and localisation. In
Fig.~\ref{fig:nt4us} (left) I show the participation ratio of the 
lowest mode as a function of the lattice gauge coupling $\beta$ for
several volumes. While below the critical coupling $\beta_c$ the PR
does not change much with the volume, above $\beta_c$ it keeps
decreasing as the system size is increased, meaning that the lowest
mode is delocalised in the low-temperature phase and localised in the 
high-temperature phase.  

Another interesting model is the SU(3) pure gauge theory, which has a
first order deconfining transition, in correspondence of which the
valence quark condensate shows a transition as well~\cite{su3}. The
authors of Ref.~\cite{KoVi} have determined the mobility edge as a
function of the gauge coupling, and they have found that it vanishes
at a value compatible with the critical $\beta$ [see
Fig.~\ref{fig:nt4us} (right)]. Again, localised modes appear at 
deconfinement.   

It is worth mentioning that the same behaviour was found in
Ref.~\cite{GKP3} in the toy model for QCD of
Ref.~\cite{GKP2}. Localised modes were found also in the
high-temperature phase in SU(2) pure gauge theory~\cite{KP}, although a
detailed study of the mobility edge as a function of temperature was
not performed.   

\section{Topology}
\label{sec:topo}

In this section I briefly discuss the possible role played by
topology in the localisation of low modes. 
The authors of Ref.~\cite{GGO} suggested that localised modes are
related to the localised zero modes associated to instantons at finite 
temperature. However, it was found in Ref.~\cite{KoVi} that even in
the SU(3) pure gauge theory case, where the instanton density is
higher than in QCD, they contribute for no more than 60\% at $T_c$,
and this fraction vanishes rapidly as $T$ increases. 

The authors of Ref.~\cite{Cossu} made a different proposal.
They showed that the localised low modes prefer locations with larger
action density and topological charge density, while the delocalised
higher modes show no particular preference. Moreover, they suggested
that the localisation might take place in correspondence with $L$ and
$\bar{L}$-type monopole-instanton pairs. On the one hand, this would
perfectly match the best ``islands'' identified by the Polyakov-line
argument, which the localised modes indeed seem to prefer. On the
other hand, these topological objects could be related to
confinement, as opposed to instantons. In any case, the role of
topology is far from being understood, and more studies are certainly
required.  

\section{Localisation in QCD-like theories in 2+1D}
\label{sec:qcdlike2+1}

In this section I discuss QCD-like theories in 2+1 dimensions. Based
on our general expectations about the relation between confining
properties and the Dirac spectrum, dimensionality should play no
particular role, as long as it is possible to have a deconfined phase
of the theory, and as long as a localisation/delocalisation transition
is possible. These theories should therefore show a close connection
between deconfinement, chiral transition and localisation of the low
modes.

However, quite a few things are different from the 3+1D case, most
notably the absence of chiral symmetry in three dimensions. Of course
one could simply check for a nonzero spectral density at the origin
without bothering about symmetries, but, as a matter of fact, for an
even number $N_f$ of flavours one can reorganise the two-dimensional 
spinors in $N_f/2$ four-dimensional spinors, for which a $U(N_f)$
chiral symmetry can be defined~\cite{Pisarski}. One can then
meaningfully ask whether this symmetry is spontaneously broken down to
$U(N_f)\to U(N_f/2)\times U(N_f/2)$ by the formation of a
fermion-antifermion condensate, and, if so, if this happens at the
onset of confinement. Given the doubling phenomenon, in this context
one type of staggered fermions counts as two flavours. 

Quite independently of this, one can also check whether localised
modes appear at deconfinement. The change in dimensionality adds a
twist to the issue, since two spatial dimensions are special for
Anderson transitions, with the existence and the nature of the
transition depending heavily on the details of the model.  

In the remainder of this section I discuss some preliminary numerical
results. 

\subsection{SU(3) pure gauge theory}
\label{sec:su3_2+1}

\begin{figure}[t]
  \centering
  \includegraphics[width=0.495\textwidth]{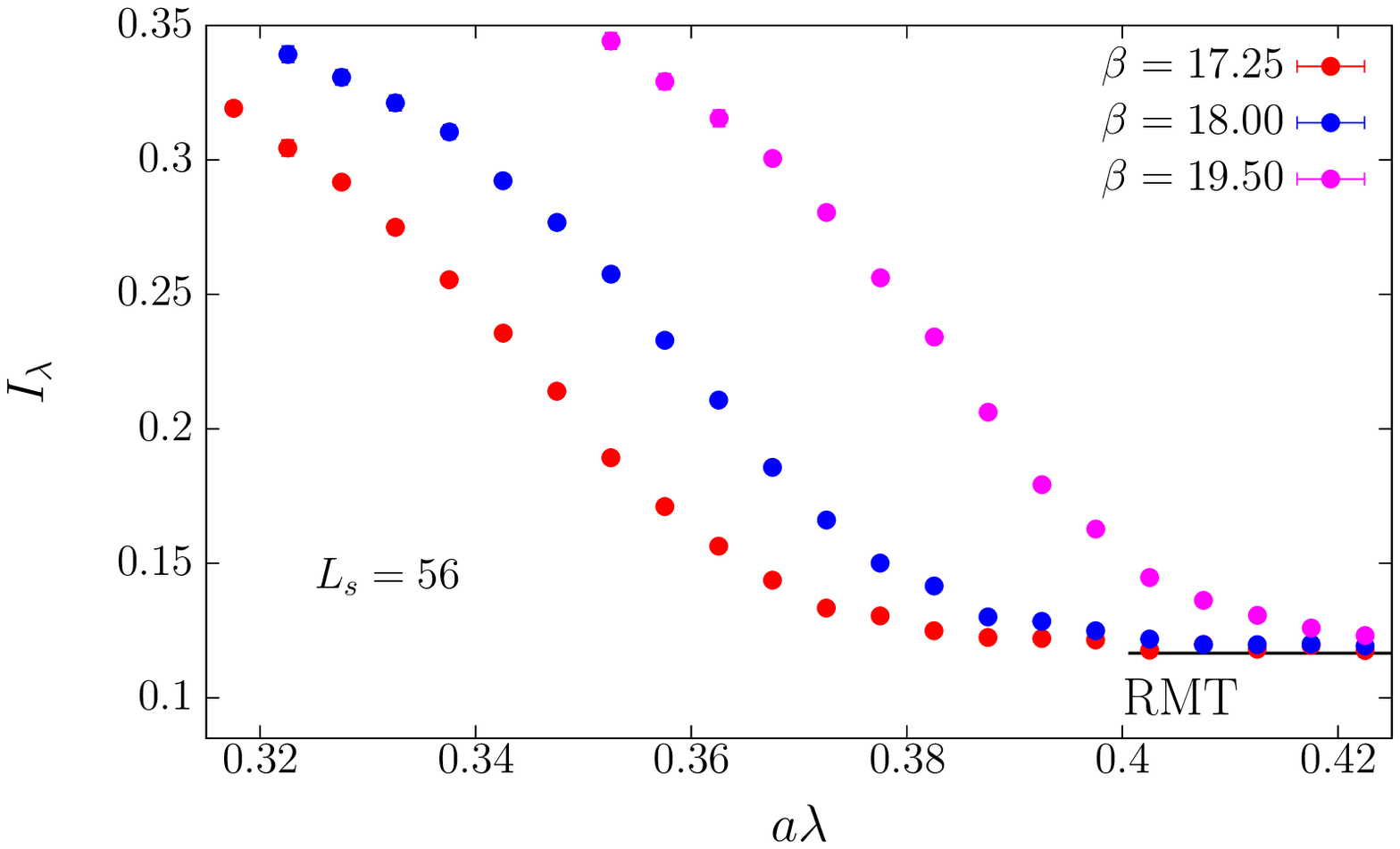}
\hspace{\stretch{1}}  \includegraphics[width=0.4925\textwidth]{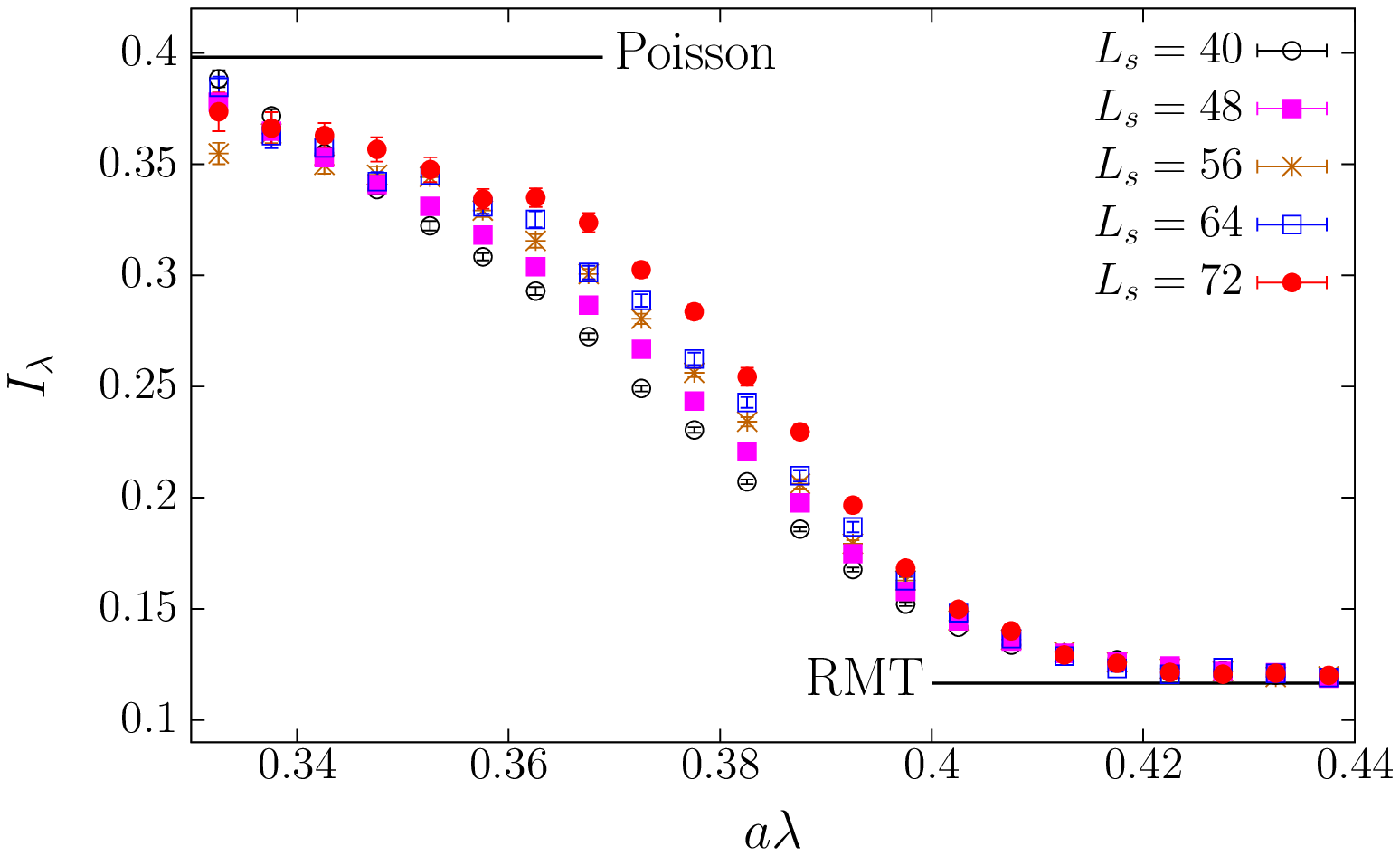}
  \caption{(Left.) Transition of $I_\lambda$ from the localised to the
    delocalised regime of the spectrum in the 2+1D SU(3) pure gauge
     theory for various $\beta$. (Right.) Same as above, but for
    fixed $\beta=19.5$ and various volumes.} 
  \label{fig:su3_2+1}
\end{figure}
As a first example I consider SU(3) pure gauge theory in 2+1D, 
studying the spectrum of the staggered Dirac operator. Deconfinement 
and chiral restoration have been shown long ago to take place at
the same critical coupling $\beta_c\simeq 15$~\cite{DHKM}. Preliminary
results show no trace of localised modes deep in the confined phase,
while localised modes are present at the low end of the spectrum in the
deconfined phase, with a transition to delocalised modes at some
critical point in the spectrum, which is qualitatively seen to
decrease with temperature [see Fig.~\ref{fig:su3_2+1} (left)]. 

An interesting aspect of the transition is that the curves of the
spectral statistics for different volumes, instead of crossing at one
point, seem to merge beyond some point in the spectrum [see
Fig.~\ref{fig:su3_2+1} (right)]. This is typical of BKT-type Anderson
transitions, and the same behaviour is observed in the analogous 2D
AM~\cite{XWL}. Defining $\lambda_c$ as the merging point,
one sees that it decreases with $\beta$. The expectation is that it
will vanish at $\beta_c$, although this cannot be confirmed at this stage.

\subsection{SU(3) pure gauge theory with imaginary chemical potential}
\label{sec:su32+1_chempot}

\begin{figure}[t]
  \centering
  \includegraphics[width=0.49\textwidth]{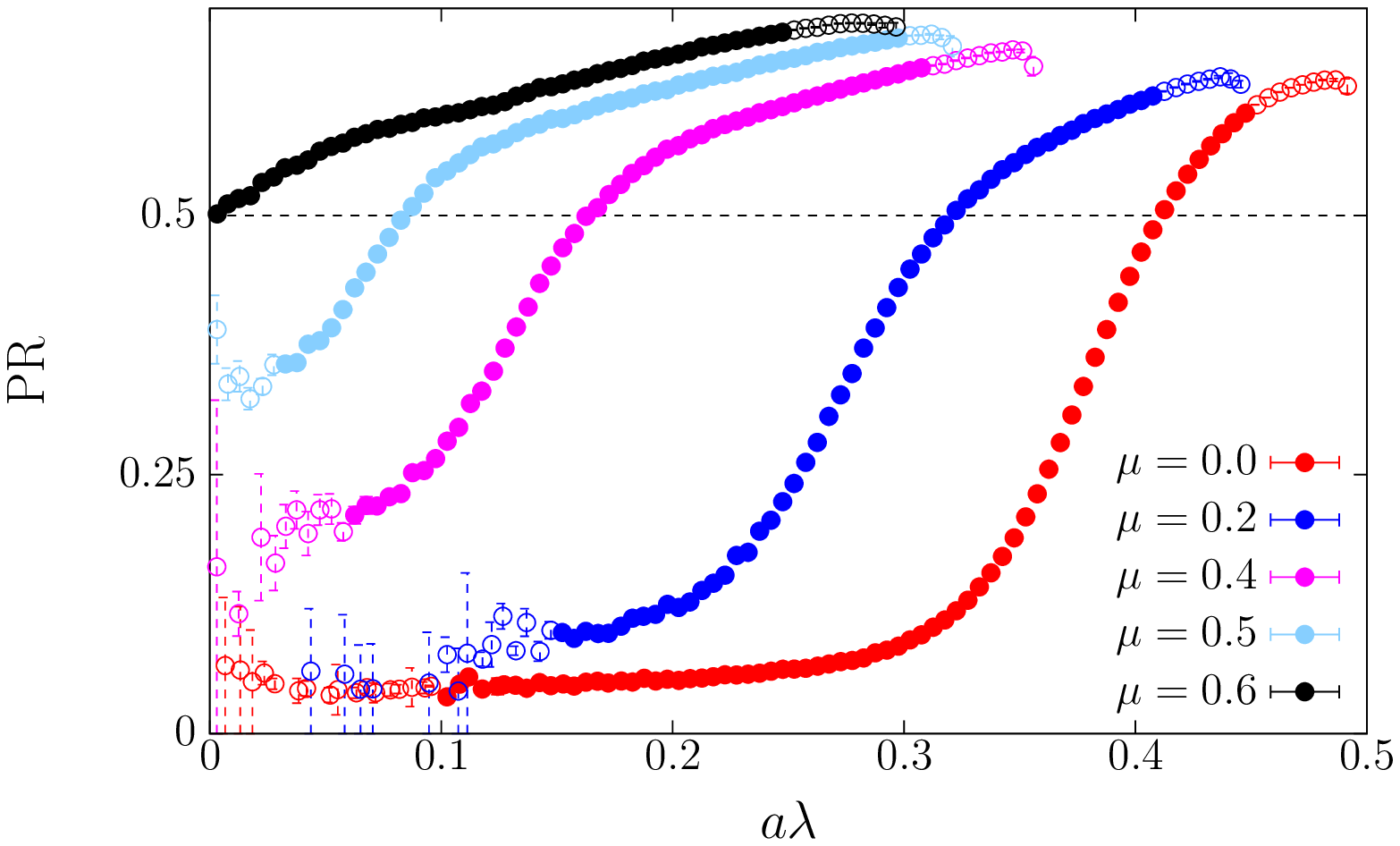}
\hspace{\stretch{1}}  \includegraphics[width=0.46\textwidth]{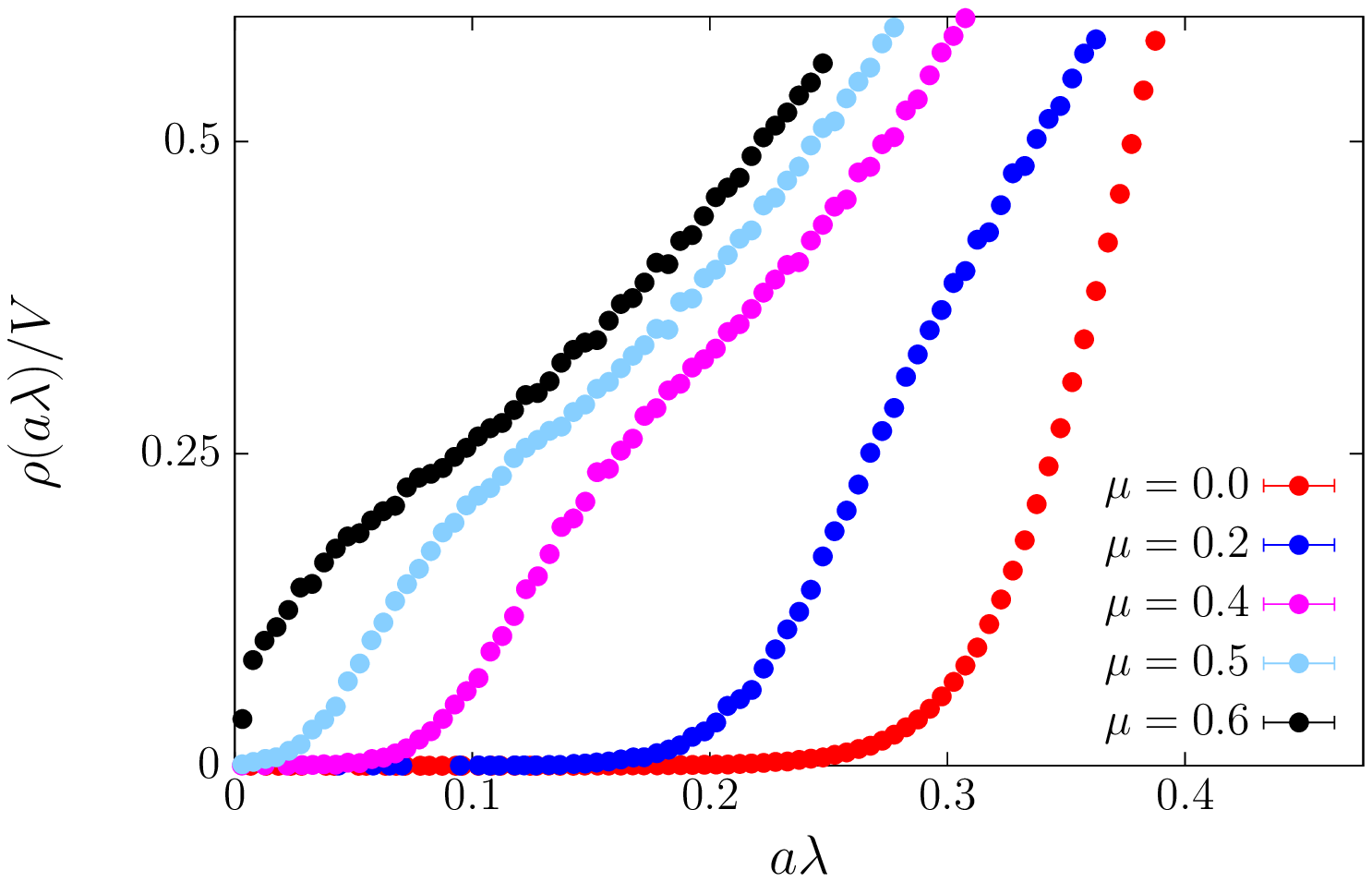}
  \caption{PR (left) and spectral density (right) in the 2+1D SU(3) pure
    gauge theory with an imaginary chemical potential.} 
  \label{fig:su3_2+1_immu}
\end{figure}

If one changes the typical value of the Polyakov line phases or the
temporal boundary conditions, one changes the typical effective
Matsubara frequencies, and thus the effective gap in the ordered
phase. According to the ``sea/islands'' picture, this should affect
the spectral density and the localisation properties of the low  modes.

One way to test this is to study SU(3) pure gauge theory in
2+1D including an imaginary chemical potential $\mu$ in the Dirac
operator, which in practice changes the temporal boundary
conditions. According to the ``sea/islands'' 
picture, increasing $\mu$ should decrease the effective gap and thus
push towards larger spectral density near the origin, and moreover
should make localisation harder. In Fig.~\ref{fig:su3_2+1_immu} (left)
I show the PR throughout the spectrum: it is clear that it increases
as $\mu$ is increased, so that the  mobility edge is pushed down,
until modes become delocalised. In Fig.~\ref{fig:su3_2+1_immu} (right)
I show the spectral density, which is indeed seen to increase with $\mu$. 

\subsection{U(1) pure gauge theory}
\label{sec:U1_2+1}

\begin{figure}[t]
  \centering
  \includegraphics[width=0.52\textwidth]{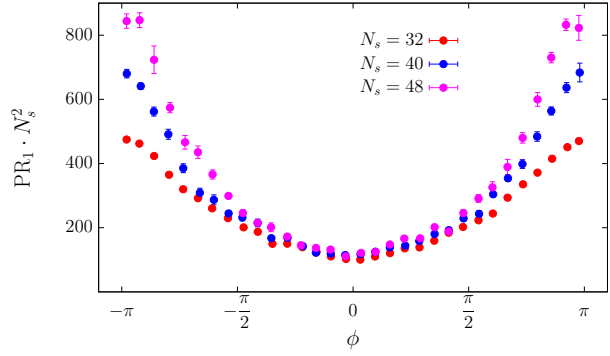}
  \caption{Average PR of the lowest eigenmode on U(1) gauge configurations with
    ${\rm arg}~\bar{P}= \phi$ for various volumes (see text).} 
  \label{fig:u1_2+1_immu}
\end{figure}

Another interesting case is 2+1D U(1) pure gauge theory,
which lacks another feature of QCD that should not matter for the
connection between deconfinement, chiral restoration and localisation,
namely the non-Abelian nature of the gauge group. Here in the
deconfined phase the absolute value of the Polyakov loop develops a
nonzero expectation value, while its phase fluctuates from
configuration to configuration, and therefore so does the effective
gap. Let $\bar{P}= (1/V)\sum_n P(n) = r e^{i\phi}$ be the spatially-averaged
Polyakov loop on a given configuration. Studying the localisation
properties of the lowest mode for different values of the phase
$\phi$, one finds that it is localised for $\phi$ near zero, while it
is delocalised for $\phi$ away from zero. This is also in agreement
with the ``sea/islands'' picture.

\section{Conclusions and outlook}
\label{sec:concl}

Deconfinement, chiral symmetry restoration and localisation of the low
modes of the Dirac operator are three closely connected phenomena,
taking place at the same temperature in QCD and similar
theories. The ``sea/islands'' picture and the Dirac-Anderson approach 
discussed in Section~\ref{sec:LQCD}, together with numerical results
in a quite diverse variety of models, support the idea that
deconfinement is the driving force behind the other two phenomena.

Insights into these issues could come from the study of other
QCD-related models, besides the 2+1 dimensional ones discussed in
Section~\ref{sec:qcdlike2+1}. In particular, the effect of an
imaginary chemical potential should be studied in the fully dynamical 
case in 3+1D. Moreover, the possible role of localisation should be
studied in the SU(3) gauge theory with adjoint fermions, where
deconfinement and chiral symmetry restoration take place at different 
temperatures. Finally, the possible role of topology should be
investigated in detail.

\section*{Acknowledgements}
This work was partly supported by grants OTKA-K-113034 and
NKFIH-KKP126769.

\end{document}